\documentclass[10pt]{article}

\usepackage{amsmath}
\usepackage{amssymb}

\usepackage[pdftex]{graphicx}
\usepackage[small]{subfigure}
\usepackage{setspace}

\usepackage{cite}
\usepackage{color}

\usepackage[labelfont=bf,labelsep=period,justification=raggedright]{caption}

\makeatletter
\renewcommand{\@biblabel}[1]{\quad#1.}
\makeatother

\date{}

\begin{document}

\begin{flushleft}
\begin{spacing}{1.5}
{\Large\textbf{Analysing Mood Patterns in the United Kingdom through Twitter Content}}
\end{spacing}
Vasileios Lampos$^{1,2,\ast}$,
Thomas Lansdall-Welfare$^{2}$,
Ricardo Araya$^{3}$,
Nello Cristianini$^{2}$\\
\bigskip
\bf{1} Department of Computer Science, University of Bristol, Bristol, UK\\
\medskip
\bf{2} Department of Computer Science, University of Sheffield, Sheffield, UK\\
\medskip
\bf{3} School of Social and Community Medicine, University of Bristol, Bristol, UK\\
\medskip
$\ast$ E-mail: v.lampos@sheffield.ac.uk
\end{flushleft}

\section*{Abstract}
Social Media offer a vast amount of geo-located and time-stamped textual content directly generated by people. This information can be analysed to obtain insights about the general state of a large population of users and to address scientific questions from a diversity of disciplines. In this work, we estimate temporal patterns of mood variation through the use of emotionally loaded words contained in Twitter messages, possibly reflecting underlying circadian and seasonal rhythms in the mood of the users. We present a method for computing mood scores from text using affective word taxonomies, and apply it to millions of tweets collected in the United Kingdom during the seasons of summer and winter. Our analysis results in the detection of strong and statistically significant circadian patterns for all the investigated mood types. Seasonal variation does not seem to register any important divergence in the signals, but a periodic oscillation within a 24-hour period is identified for each mood type. The main common characteristic for all emotions is their mid-morning peak, however their mood score patterns differ in the evenings.


\section*{Introduction}
Social Media, particularly the micro-blogging website Twitter, provide a novel way to gather real time data in large quantities directly from users. This data, which is also time-stamped and geo-located, can be analysed in various ways to examine patterns in a wide range of subjects. Several methods have been already proposed for exploiting this rich information in order to detect events emerging in populations \cite{Lampos2011a}, track possible epidemics \cite{Lampos2010f,Lampos2010,Signorini2011}, model opinion polls \cite{OConnor2010,Lampos2013} or even infer results of elections \cite{tumasjan2010predicting}.

People's mood changes throughout the day, spontaneously or in response to life's vicissitudes. Exploring these variations in real time and with large samples is difficult, often relying on self-reporting. The use of information gathered via Twitter messaging may provide a method to overcome some of these difficulties, allowing for the assessment of mood states in a set of users through analysing the contents of their textual communications in real time and on a large scale. A better knowledge of patterns of mood variations would be useful to explore widely accepted notions linking certain psychiatric phenomena to temporal -- or potentially even geographical -- patterns \cite{Kronfeld-Schor2012,Grimaldi2009}. In this study we focus on the possibility to reliably measure a temporal mood signal from a vast quantity of data.

Approximately 15\% of the overall adult population in the United Kingdom (UK) uses Twitter, with a small bias towards young male adult users (56\%) \cite{IPSOSMediaCT2012}. Furthermore, 61\% of Twitter users are regarded as of upper middle, lower middle or middle class \cite{IPSOSMediaCT2012}. Hence, any findings need to be considered as representative of this sample of the general population.

Recent studies using information captured from Twitter messaging found circadian and seasonal patterns in the content of emotionally loaded wording \cite{Golder2011,Lansdall-Welfare2012}. We used a similar methodology to analyse data we have gathered from the UK in two distinct time intervals of 12 weeks each, forming two textual time series. The analysis produces a numeric time series for each mood indicator, which is later subjected to statistical analysis. This paper aims to establish whether the mood patterns ascertained via Twitter messaging follow periodic time patterns, especially in terms of diurnal and seasonal variations.

\section*{Analysis}
We collected approximately 120 million tweets in two different time intervals of 12 weeks each, 50 million in the winter 2011 and 70 million in the summer 2011, from the 54 most populated urban centres in the UK. This was done by periodically retrieving the 100 most recent tweets, geo-located to within a 10km range of an urban centre in our list. Geo-location is based on Twitter's service for users, who have opted-in. Twitter tracks their location via their IP address or mobile service. We attempted to reduce any potential sampling bias by applying the same daily sampling frequency in each urban centre. Data was collected for 12 weeks during each season (06/12/2010 to 28/02/2011 for winter and 06/06/2011 to 28/08/2011 for summer). Our choice of collecting tweets geo-located in urban centres was justified by the fact that in rural areas the ratio of messages over the population is expected to be significantly lower \cite{Mislove2011}.

We obtained a stemmed version of the text by applying Porter's Algorithm, which performs a homogeneous calibration on the suffix of a word (e.g. `happy' is converted to `happi', but the same stands for `happiness') \cite{porter1980}. This is a standard practice in text analysis. The tweets were divided into 24 bins, one for each hour of the day, and the emotional valence of each hour was assessed by a text analysis tool, described below.

We measure the emotional valence of a text by counting the frequency of certain emotionally loaded words. These words are provided by WordNet Affect, a freely available resource that provides an emotional score for each English word \cite{Strapparava2004}. We estimated the level of activity for four emotions - namely fear, sadness, joy and anger - by counting the frequency of these mood-related terms within the content of Twitter messages. WordNet Affect words were also stemmed. After this pre-processing, 146 word-stems were found to represent anger, 92 fear (the word `alarm' was removed from the word list for fear, since it showed a strong linear correlation -- summer: $0.877$ and winter: $0.866$, $p < 0.001$ for both - with the occurrence of the word `clock', resulting in a peak when people wake up), 224 joy and 115 sadness. WordNet Affect as well as stemming have been regularly applied in other emotion-detection and text mining applications and thus, are considered as standard ways to retrieve information of this type \cite{Strapparava2008,Calvo2010,Acerbi2013}.

\subsection*{Mood Score Computation}
The central piece of technology necessary for this study is a function that can map text documents to numeric values reflecting the prevalence of a certain mood-type (mood score). For each mood type there is a list of associated terms (derived from WordNet Affect). The score is computed as follows. For each mood-term in the list, its relative frequency is computed for each hourly interval (i.e., by dividing its count by the total number of words observed in that interval). This time series is then standardised, subtracting its mean and dividing it by its standard deviation. The score of each of the mood-types is obtained by averaging this quantity (standardised relative term frequency) over all terms associated to that mood-type. Standard errors (SEs) of the mean mood-scores were incorporated in all the plotted time series using fading colours. SEs were calculated by dividing the standard deviation of all observations for that interval by the square root of the number of observations.

\subsection*{Testing Circadian Patterns (TCP)}
All the resulting time series were tested for the presence of a significant periodic pattern of 24 hours. A time series shows a circadian pattern, if it shows a periodic structure within a period of 24 hours. The statistical significance of a circadian pattern was assessed by performing the following test (TCP). The pair-wise mood-score Pearson correlation coefficients were computed for all the daily time series (168 samples) in the data set and mean pairwise correlations were derived ($c$). Then, the time series for each day was randomised and the mean pairwise correlation for this randomised data was computed ($r$). By performing the latter $n$ times and counting the number of times -- say k --, where $r \geq c$, we computed the p-value $p = \frac{k}{n}$. This p-value estimates the probability that a set of randomised time series has a greater average pairwise correlation than the one computed in the original (non-randomised) data set.  As a further method to objectively detect a periodic structure in the time series, we have included an autocorrelation analysis of the mood signals. Autocorrelation simply depicts the Pearson correlation of a mood signal with a lagged version of itself and is a basic method for investigating the existence of periodicities in the signal together with the corresponding frequencies.

\subsection*{Testing Mood-Score Differences (TMD) and Peaking Times (TPT)}
We have also considered whether there exist statistically significant differences between specific time periods of interest (\emph{e.g.}, morning hours versus evening hours) as well as between seasons (\emph{i.e.} morning hours in winter versus summer) and peaking moments in the mood scores. For testing differences in mood-scores (TMD), we estimated average time series using bootstrap sampling with replacement \cite{Efron1993}. In each bootstrap (we ran 10,000 bootstraps per mood type), the seasonal daily mood data are sampled with replacement (65\% unique daily samples per bootstrap) and an average time series is computed. When comparing two sets of hourly intervals $H_a$ and $H_b$, the p-value for the hypothesis that the mood score in $H_a$ is greater than in $H_b$ indicates the number of times that $H_b$'s mood scores were greater or equal to $H_a$'s, divided by the total number of bootstrap samples involved. Similarly, for the hypothesis that a mood score has a peak (or a trough) during a set of hourly intervals $H_c$, the p-value represents the times that the mood score did not reach a maximum (or respectively a minimum) during $H_c$ divided by the number of bootstrap samples used (TPT).

\section*{Results}
\begin{figure}[t]
\begin{center}
\includegraphics[width=0.9\textwidth]{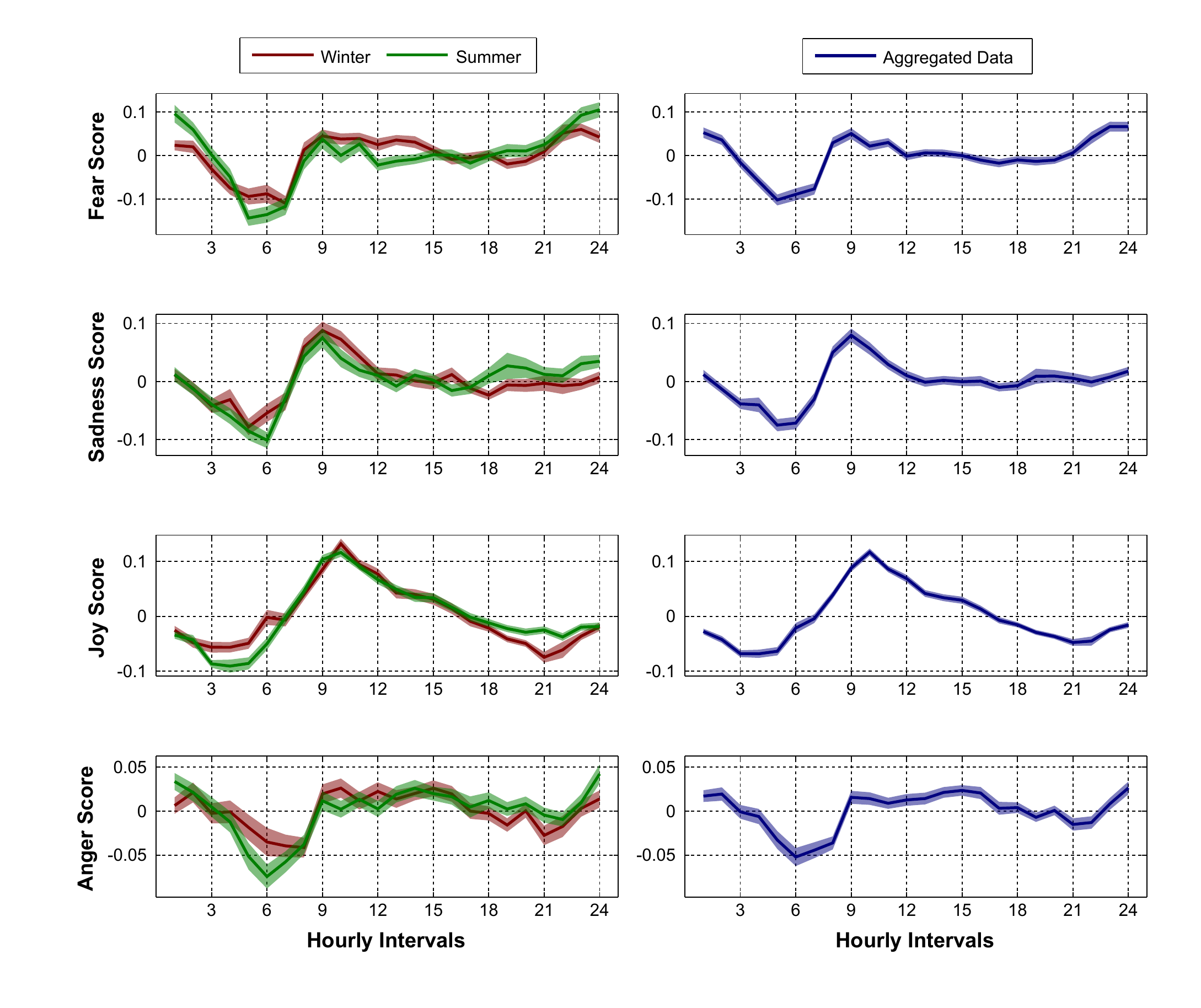}
\end{center}
\caption{
{\bf Plots representing the variation over a 24-hour period of the emotional valence for fear, sadness, joy and anger.} The red line represents days in the winter, while the green one represents days in the summer. The average circadian pattern was extracted by aggregating the two seasonal data sets. Faded colourings represent the SE of the sample mean.
}
\label{Figure_1}
\end{figure}

We found that all time series corresponding to mood types show a strong periodicity of 24 hours, and the average 24-hour profile (circadian pattern) is reported in Figure \ref{Figure_1}. The 24-hour periodicity in the mood patterns was statistically significant for all mood types ($p_{\text{TCP}} < 0.001$).  For each emotional type, we have computed the circadian pattern for winter and summer as well as an aggregation of those data sets (we also refer to this pattern as average). Figure \ref{Figure_1} also shows a breakdown by mood-type and by season of data collection. There were no clear differences in these patterns across seasons.

As far as the specific patterns for emotions are concerned, they all seem to peak at around 9a.m. after an understandable decline throughout the night. However there seems to be substantial variation after peak times and especially at evenings. For instance fear shows a moderate decline until 8p.m. (8a.m. -- 8p.m., $p_{\text{TMD}} = 0.032$) when it starts increasing again until midnight (8p.m. -- 12a.m., $p_{\text{TMD}} < 0.001$).  Sadness decreases from the morning to 1p.m. (9a.m. -- 1p.m., $p_{\text{TMD}} < 0.001$) after which it becomes steady for the rest of the day. Joy decreases and remains low until 5a.m. (9a.m. - 5a.m., $p_{\text{TMD}} < 0.001$). The lowest scores for joy happen either early or very late at night (8 to 10p.m. or 2 to 5a.m., $p_{\text{TPT}} = 0.003$). Anger is quite stable after 9a.m., but shows a small decrease in the evening.

\begin{figure}[t]
\begin{center}
    \subfigure[Fear]{\includegraphics[width=0.47\textwidth]{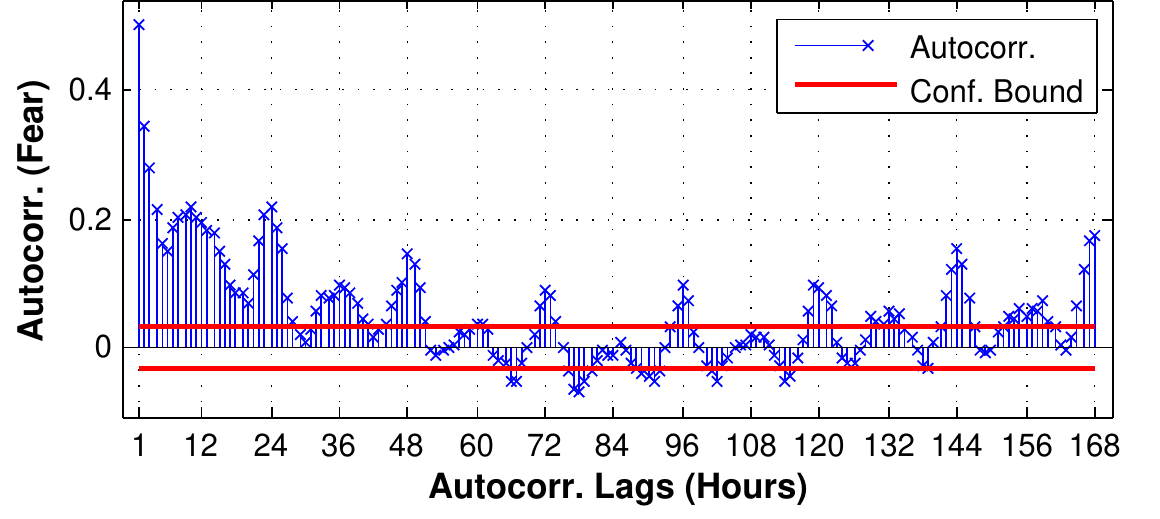}
    \label{figure_2a}}
    \subfigure[Sadness]{\includegraphics[width=0.47\textwidth]{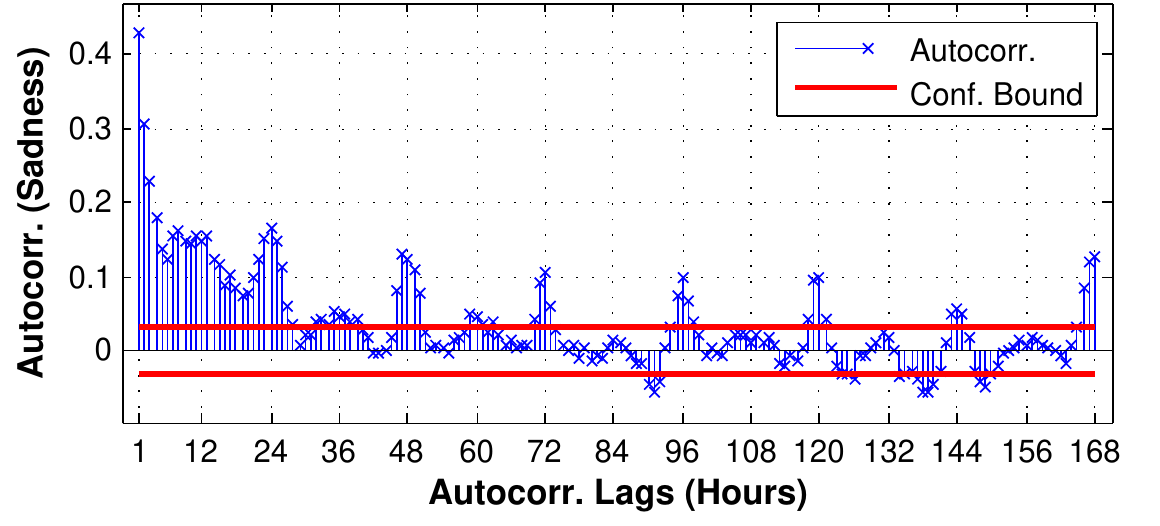}
    \label{figure_2b}}
    
    \quad
    
    \subfigure[Joy]{\includegraphics[width=0.47\textwidth]{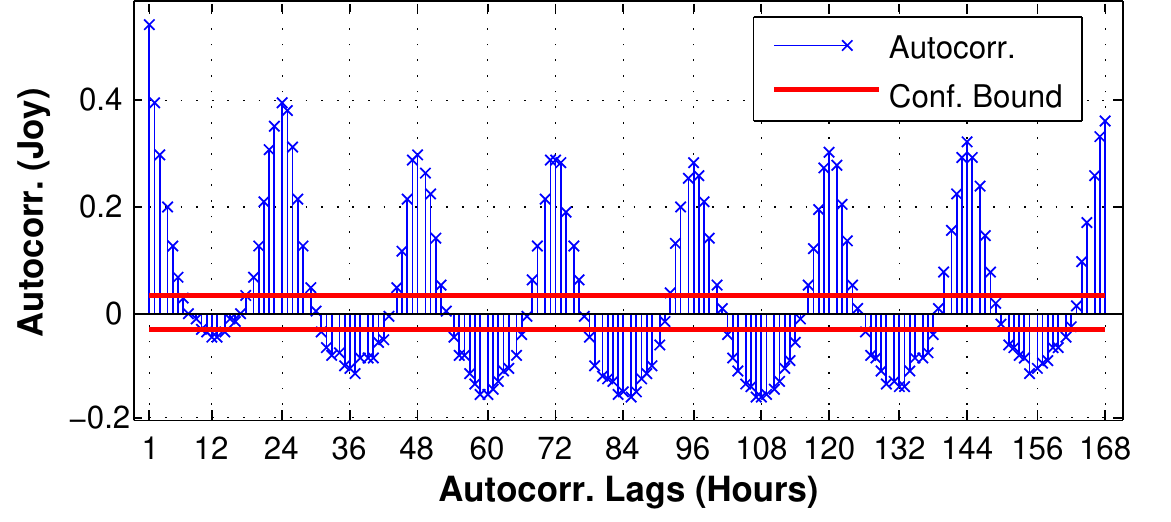}
    \label{figure_2c}}
    \subfigure[Anger]{\includegraphics[width=0.47\textwidth]{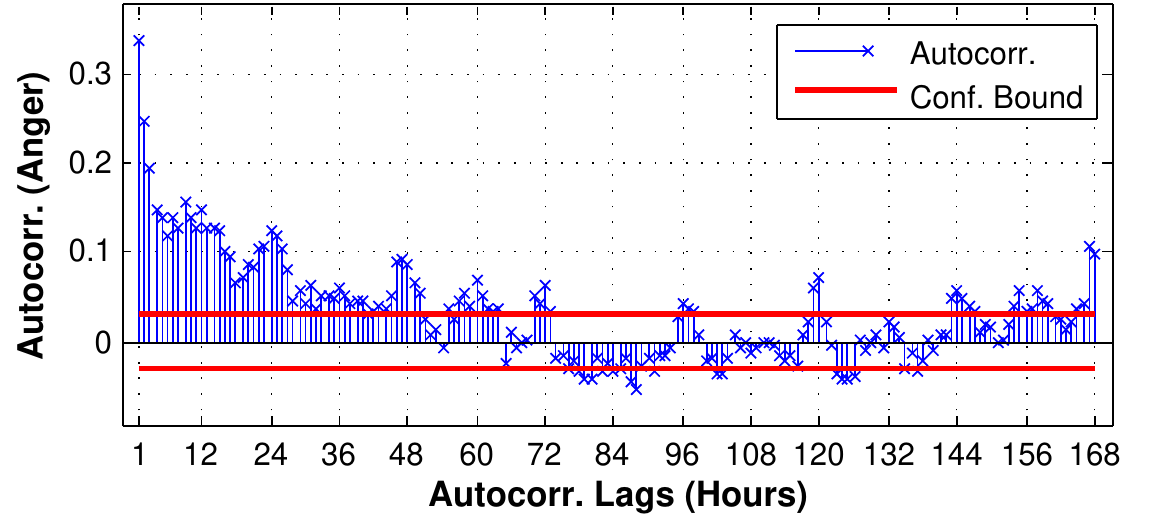}
    \label{figure_2d}}
\end{center}
\caption{
{\bf Autocorrelation figures for
the mood patters of fear, sadness, joy and anger.} The considered lags range from 1 to 168 hours (7 days). Correlations outside the confidence bounds are statistically significant.
}
\label{Figure_2}
\end{figure}

Autocorrelation plots (Figure \ref{Figure_2}) for the four mood types using lags ($\ell$) ranging from 1 to 168 hours, \emph{i.e.} the total number of hours in a week, show that consecutive time intervals (hours) are unsurprisingly highly correlated for all emotional types. From the autocorrelation plots, it also becomes evident that for each emotional type there exists some level of daily ($\ell = 24$) as well as weekly periodicity ($\ell = 168$). Autocorrelations for $\ell \in \{25,..., 167\}$ are lower than the one observed for $\ell = 168$ indicating that the weekly pattern (or period) in the mood signal is stronger than the intermediate ones. Overall, the emotion of joy has the highest levels of autocorrelation showing the strongest periodic behaviour, whereas periodicity seems to be less strong for the mood type of anger.

\section*{Discussion}
As far as we are aware this is the first study of real-time mood variation at a population level using Social Media information in the UK. Similar studies have been proposed for other populations, and in this respect our study is a novel contribution to the literature in the subject \cite{Golder2011,Lansdall-Welfare2012}. This study shows that it is possible to estimate aggregate mood states in a large population by analysing the contents of its communications via Social Media. We detected a strong circadian pattern for all the emotions we investigated, in keeping with previous studies \cite{Golder2011,Lansdall-Welfare2012}. However, we found that mood patterns for sadness did not agree with currently held clinical concepts of diurnal variation of mood or increased prevalence of depressive mood during the winter months. In fact, unlike for the 24-hour strong periodicity pattern for all emotions, we found no clear seasonal variations for any of the emotions studied.

\subsection*{Comparison with other relevant general population studies}
Recent studies have also exploited Twitter content for extracting circadian and seasonal patterns \cite{Golder2011,Lansdall-Welfare2012}. A study found that the volume of words representing negative affect reached their highest point early in the mornings (3-4a.m.) \cite{Golder2011}, somehow out of keeping with the clinical concept of diurnal variation of mood among depressed patients. Seasonal patterns, in particular an increase of depressive symptoms in winter months, were not confirmed either \cite{Golder2011}. In addition, this work presents mood patterns that appear to be different from the ones we have extracted, something that could be the result of merging tweets geo-located in Australia and UK as well as not using a standardised frequency of the mood terms \cite{Golder2011}. However, this study showed seasonal changes in a concept defined as Positive Affectivity (PA, that includes feelings like enthusiasm, delight, activeness, and alertness) and speculated that "winter blues" may be associated with diminished Positive Affectivity rather than increased Negative Affectivity (NA, that includes feelings like distress, fear, anger, guilt, and disgust) \cite{Golder2011}. There is further support to this hypothesis.  A small study comparing healthy and depressed individuals with repeated measurements of NA and PA found different patterns. Healthy individuals showed significantly larger drops in PA throughout the day compared to depressed individuals but NA was only present in the mornings for depressed individuals \cite{Peeters2006}. Another study with a small number of healthy women with various levels of depressive symptoms found that those with more depressive symptoms showed lower peaks of PA in the morning but a similar decline in PA throughout the day in comparison with those with less depressive symptoms \cite{Murray2007}. Interestingly, in our study we found that the scores of sadness and joy, most likely proxies of PA and NA, show similar patterns in the mornings but sadness remains reasonably high during the day, whereas joy shows a pronounced decline throughout the day, in keeping with the concept that it is PA rather than NA that shows the more visible circadian changes in the general population (note that our definition of PA and NA is not exactly equivalent to that of the studies cited above).

\subsection*{Comparison with clinical sample studies and possible clinical implications}
It has been widely believed that an increased intensity and higher load of depressive symptoms early in the mornings is a core symptom of major depression, especially of the melancholic sub-type \cite{Peeters2006,Morris2009}. Although messages with a sadness connotation peaked early in the morning in our study, these patterns seem to be non-specific because they were seen for all emotions.  Other studies have found that both positive (joy) and negative (sadness) emotions show different patterns throughout the day in healthy and depressed subjects as mentioned above \cite{Peeters2006}. There seems to be a much clearer difference in circadian patterns of PA when comparing normal and depressed individuals, with the latter showing either a steady or an increasing pattern of PA throughout the day and healthy individuals a clear decline in PA, as we found in our study (joy).  Besides it seems that the morning peak of NA is present only for depressed rather than healthy individuals \cite{Peeters2006}, suggesting that this may be a specific clinical feature of depression.

Our study shows clearly that mood changes throughout the day in consistent patterns. Exploring these variations in real time and with large samples is difficult to accomplish. The use of information gathered via Twitter messaging permits overcoming some of these difficulties. A better knowledge of patterns of mood variations would be useful to explore widely accepted notions linking certain psychiatric phenomena to diurnal or even seasonal patterns \cite{Kronfeld-Schor2012,Grimaldi2009}. Interestingly, we found no evidence to support seasonal mood variations in this population of Twitter users. Nonetheless, it is important to emphasize once again that clinical samples may differ substantially from specific general populations such as the one studied here. In the future we may be in the position to record in real-time mood state variations among depressed patients using newly emerging technological devices such as smartphones and biosensors. It is also essential to highlight that contextual factors could influence both the mood experienced and the reporting of this data. A major upset may lead to sadness which may be reported or not depending on a number of variables such as willingness or availability of means. It is also possible that Twitter messaging may induce or confirm emotional states, thus increasing the volume of message transit of such emotions. Despite all this, the massive aggregated data derived from Twitter allows mapping out overall patterns of mood in the population of users, which is consistently growing.

\subsection*{Limitations}
There are limitations when conducting these studies. Among these, it is obvious that people cannot send messages whilst asleep and therefore there is a drop in signalling over night. However, we adjusted our results by both normalising and standardising to reduce the impact of this effect. In this work, we did not consider the influence that significant events or natural phenomena emerging in real-life might have on the extracted signals and our method for collecting tweets excluded content geo-located in rural UK areas. Nonetheless, we argue that such biases are usually resolved to a significant level, when working with large-scale samples of data; the periodical characteristics and the statistical significance provide further proof for the stability of the extracted signals. Our findings apply only to the population of Twitter users geo-located in the UK. The analysis cannot disentangle within-person from between-person effects; hence, the results represent a collective behaviour. Limitations arising due to attributes of the general population that might not be present in Twitter users, such as behavioural motives of the elderly, do create unresolved biases; therefore, in this work, we can only make claims about the population of UK Twitter users and not the general population. Further work is also needed to separate different types of users, so that important information is not lost by means of averaging.

WordNet Affect itself is not an exhaustive resource of affective terms and does not quantify the relative contribution (weight) of each term. Moreover, some terms related to mood may take different meanings based on the context. This is an obvious limitation of bag-of-words methodologies like ours, which consider unigrams (single words) as independent sources of information. Further validation of the affective words using, for example, labelled data and supervised statistical learning techniques might improve the accuracy of the proposed mood scoring schemes.

In spite of all this, Twitter does represent a great opportunity to capture real-time data for massive numbers of people in the general population and use it to understand several aspects of human behaviour and life. For the first time we can analyse the contents of communications among huge masses of people, and this is likely to contain precious information for social scientists, anthropologists and - no doubt - psychiatrists. Further research will be carried out in this direction, to include geography, time patterns, and segmentation of different sub-populations, for example by age or gender. In the future it will be possible to trace more accurately individuals as they move about and capture layers of information that can allow us to study variations within as well as between persons. Various technologies and data capturing methods are being integrated to get an improved understanding of how people feel and behave. Of course, the availability of better technology to pursue these endeavours brings also huge ethical dilemmas in terms of how to protect data confidentiality and privacy.

\section*{Acknowledgments}
This work has been supported by the EU-FP7 project `COMPLACS' and the PASCAL2 Network of Excellence.

\bibliography{../../../../Bibtex/library}

\end{document}